\documentclass[
 aps,
 reprint,
 prb,
 superscriptaddress
]{revtex4-2}

\usepackage[utf8]{inputenc}
\usepackage[a4paper,width=170mm,top=25mm,bottom=30mm]{geometry}
\usepackage{multirow}
\usepackage{footnote}
\usepackage{graphicx}
\usepackage{braket}
\usepackage{amsmath}
\usepackage{amssymb}
\usepackage{booktabs}
\usepackage{upgreek}
\usepackage{subfigure}
\usepackage[backref=none]{hyperref}
\hypersetup{
    colorlinks=true,
    linkcolor=blue,
    citecolor=blue,
    urlcolor=red,
}
\usepackage{enumitem}
\usepackage{esint}
\usepackage{lipsum}

\newcommand{\pdag}{\phantom{\dagger}}
\newcommand{\vk}{\mathbf{k}}

\begin{document}
\title{Reduced pair breaking from extended disorder in unconventional superconductors: implications for 4Hb-TaS$_2$}
\author{Yuval Tsur}
\affiliation{Department of Physics, Bar-Ilan University, 52900, Ramat Gan, Israel}

\author{Mark H Fischer}
\affiliation{Department of Physics, University of Zurich, Zürich, Switzerland}

\author{Jonathan Ruhman}
\affiliation{Department of Physics, Bar-Ilan University, 52900, Ramat Gan, Israel}

\date{\today}
\begin{abstract}
Unconventional superconductivity is generally expected to be strongly suppressed by nonmagnetic disorder, as captured by Abrikosov--Gor'kov (AG) theory. However, several materials, including transition metal dichalcogenides, exhibit signatures of unconventional pairing despite relatively high resistivities, suggesting a breakdown of the conventional relation between momentum relaxation and pair breaking. 
Here, we study this problem in H-phase transition metal dichalcogenides by computing the disorder-dressed pairing susceptibility. We employ a multiband model with spin-orbit coupling and include an impurity potential that mimics a common lattice defect, namely a chalcogen vacancy or site ad-atom. 
This yields to an extended impurity potential, which we compare with the commonly considered on-site (point defect) potential. 
We evaluate  the momentum-relaxation rate and the pair-breaking rate on equal footing.
We find that extended impurity potentials lead to a parametrically reduced pair-breaking rate compared to the transport scattering rate, with $\Gamma \tau_D \sim 1/3$ over a wide parameter range. This reduction originates from the momentum structure of the disorder potential, which partially matches the internal structure of the superconducting gap and suppresses pair-breaking processes. As a result, unconventional pairing states are significantly more robust than predicted by standard AG theory.
Our results provide a natural explanation for the persistence of unconventional superconductivity in systems with strong disorder and substantially alleviate the apparent conflict between high resistivity and unconventional pairing in materials such as 4Hb-TaS$_2$.
\end{abstract}

\maketitle

\section{Introduction}
The study of unconventional superconductivity~\cite{norman2011challenge} exposes rich emergent low-energy phenomena, ranging from novel collective bosonic modes to exotic fermionic and potentially anyonic excitations. Despite this richness, much of this phenomenological landscape remains only partially explored, largely due to the challenge of unambiguously identifying unconventional superconducting states. One of the standard identifying properties of unconventional pairing, namely their response to disorder, has recently been under renewed scrutiny.

The celebrated theory of Abrikosov and Gor'kov (AG)~\cite{AAAbrikosov1959} predicts that non-$s$-wave superconductors are highly sensitive even to non-magnetic disorder. Within the AG framework, superconductivity is destroyed once the mean free path $l$ becomes comparable to the superconducting coherence length $\xi$, or equivalently when the single-particle scattering rate reaches the scale of the clean transition temperature, $\hbar/\tau \sim k_{\rm B}T_{\rm c}^{(0)}$. This conclusion relies on several assumptions: isotropic impurity scattering treated within the (self-consistent) Born approximation, negligible vertex corrections, and an effective single-band description, in which the momentum relaxation rate directly sets the pair-breaking rate. Under these conditions, the momentum relaxation rate $1/\tau$, which can be estimated from transport measurements, provides a direct measure of the pair-breaking strength.

However, a growing number of candidate non-$s$-wave superconductors appear to violate the AG expectation. Materials, such as the cuprates~\cite{juskus2024insensitivity}, doped topological insulators~\cite{hor2010BiSe,Novak2013SnTe}, iron-based superconductors~\cite{przorov2014effect,Teknowijoyo2026FeSe}, transition-metal dichalcogenides~\cite{fischer:2023,simon2024transition}, and certain kagome superconductors~\cite{holbaek_2023} exhibit signatures of unconventional pairing despite disorder levels that would naively imply $\hbar/\tau \gtrsim k_{\rm B}T_{\rm c}$. However, resistivity probes the momentum relaxation rate, whereas superconductivity is controlled by the pair-breaking rate $\Gamma$.  These need not be equivalent: For example, in multiband systems with internal anisotropy, pair breaking is governed primarily by interband scattering processes that mix gap sectors with opposite signs, while intraband scattering dominates momentum relaxation~\cite{hirschfeld2011gap}. Consequently, a large resistivity does not necessarily imply strong suppression of unconventional pairing. More generally, pair breaking is sensitive to scattering between regions with different gap phases, whereas the transport lifetime is controlled by backscattering, and the two need not coincide.

In this context, the transition-metal dichalcogenide 4Hb-TaS$_2$ provides an intriguing example. It exhibits signatures of unconventional superconductivity together with a relatively high resistivity~\cite{Ribak2020}. Structurally, 4Hb-TaS$_2$ consists of alternating H (trigonal prismatic) and T (rhombohedral) layers held together by van der Waals forces. The bulk T phase is a correlated insulator proposed to realize a quantum spin liquid~\cite{Law2017,Ribak2017}, whereas the bulk H phase is metallic and superconducting at low temperatures. Unlike bulk 2H-TaS$_2$, however, the 4Hb polytype displays a range of unconventional phenomena, including gapless edge modes, a $\pi$-phase shift in Little--Parks experiments~\cite{Almoalem2024}, magnetic memory above $T_{\rm c}$~\cite{Persky2023}, finite thermal conductivity and specific heat far below $T_{\rm c}$~\cite{d13p-mtbz}, and an enhanced muon depolarization rate often interpreted as evidence for time-reversal-symmetry breaking~\cite{Ribak2020}. At the same time, the measured resistivity is approximately 65~$\mu\Omega$cm~\cite{fischer:2023}, which corresponds to a naive estimate of the transport life time $\hbar/\tau_{\rm tr} \sim 10$~meV, far exceeding the superconducting energy scale $k_{\rm B}T_{\rm c} \sim 0.2$~meV~\cite{Nayak2021}.

Several theoretical works have recently addressed similar discrepancies by generalizing AG theory to multiband systems or systems with strong spin-orbit coupling. In particular, Michaeli and Fu showed that odd-parity pairing in Dirac systems can be robust against density disorder~\cite{michaeli2012spin,cavanagh2020robustness}, a result later understood as a symmetry-based generalization of Anderson’s theorem~\cite{timmons2020electron} involving the product of chiral and time-reversal symmetry~\cite{Dentelski2020}. 
A comprehensive picture is obtained  when considering all symmetry-allowed inter-orbital scattering processes~\cite{cavanagh2021general}. A further route to stabilizing unconventional superconductivity incorporates the specific site symmetry of the scatterer~\cite{holbaek_2023, holbaek:2026tmp}. Additionally, related ideas have been formulated in terms of superconducting fitness~\cite{ramires2018tailoring,cavanagh2020robustness,andersen2020generalized,cavanagh2021general}. Finally, in the context of monolayer transition-metal dichalcogenides, it has been shown that Ising spin-orbit coupling can induce parity-mixed superconductivity and qualitatively modify the response to magnetic fields and scalar disorder~\cite{mockli2018robust,mockli2019magnetic,mockli2020ising}. These approaches retain the core structure of AG theory but reveal that symmetry, band structure, and spin-orbit coupling can strongly suppress effective pair breaking. An open question is whether one must go even further beyond the standard AG assumptions—for instance by incorporating vertex corrections in the impurity problem—to obtain a complete description~\footnote{The AG framework typically neglects vertex corrections in the impurity-averaged pairing interaction. In systems with strongly momentum-dependent or extended impurity potentials, such corrections may further renormalize the effective pair-breaking rate.}. 

Motivated by these developments, we investigate disorder effects in 4Hb-TaS$_2$ within an AG-based framework adapted to its realistic band structure with Ising spin-orbit coupling and study various non-$s$-wave pairing states. Treating sulfur--selenium substitution as the main source of disorder, we compare the resulting finite-range impurity potentials with conventional on-site scattering. { Our goal is to isolate the role of extended impurity potentials within a controlled microscopic model, neglecting other potentially important ingredients, such as multiorbital and multilayer effects.} 

Since experiments primarily measure resistivity rather than the microscopic scattering rate, we analyze the ratio between the computed pair-breaking rate $\Gamma$ and the momentum relaxation rate $1/\tau$ obtained from Boltzmann transport theory for fixed impurity concentration and coupling strength. We find that the dimensionless quantity $\Gamma \tau$  depends on the strength of the spin-orbit coupling~\cite{mockli2018robust,mockli2019magnetic,mockli2020ising}, but even stronger on the impurity range. For realistic parameters and extended-defect potentials, certain non-$s$-wave pairing states can withstand momentum lifetimes up to five times shorter than predicted by naive AG theory. While our results do not fully eliminate the discrepancy between experiment and conventional expectations, they substantially narrow the gap by demonstrating that large momentum relaxation rates do not necessarily imply comparably strong pair breaking.

The rest of this paper is organized as follows. In Section II, we present our model Hamiltonian, which includes an effective single-orbital model and two types of disorder potentials considered in this work: point-like and extended defects. In section III, we review AG theory and describe how we generalize it to the specific case of Ising SOC in a crystal with $D_{3h}$ symmetry. In Section IV, we present the results and discuss their implications in Section V. 

\section{The model}
The transition metal dichalcogenide 4Hb-TaS$_2$ consists of alternating layers of 1H and 1T structures with a four-layer unit cell and crystallizes in the P6$_3$/mmc space group (\#194) with point group $D_{6h}$. The electronic structure is highly two-dimensional~\cite{almoalem2024charge} with the low-energy electronic degrees of freedom mostly living on the 1H layers. Therefore,   we focus on the physics of a single H layer of TaS$_2$ hereafter. Such an individual layer has only $D_{3h}$ symmetry, which implies it lacks inversion symmetry. This lack of inversion manifests itself in a strong Ising-type spin-orbit splitting in its electronic band structure, a setting known to permit symmetry-allowed singlet--triplet mixing and unconventional superconducting responses to magnetic fields and disorder~\cite{mockli2018robust,mockli2019magnetic,mockli2020ising}.

\subsection{Effective Band Hamiltonian}\label{subsec:band_hamiltonian}
\begin{figure}
    \centering
\includegraphics{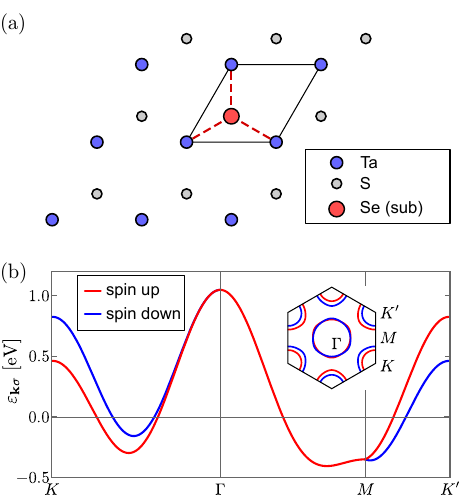}
    \caption{(a) Schematic of the 1H-TaS$_2$ lattice showing a unit cell and a defect. The black rhombus outlines the pristine unit cell. A single sulfur site within the cell is replaced by Selenium (red), and its coupling to the nearest Ta neighbors is shown with red dashed lines. (b) Band structure of 1H-TaS$_2$ of the effective model Eq.~\eqref{eq:H} along the contour $ K  - \Gamma - M - K'$ with $t_{\rm NN}=0.09$ eV, $t_{\rm NNN} = 0.26$ eV, and $\lambda/t_{\rm NN} \approx 0.8$. } 
\label{fig:H-structure}
\end{figure}

The electronic properties of 1H-TaS$_2$ are dominated by the hopping of Ta $d$ electrons on a triangular lattice, as depicted in Fig.~\ref{fig:H-structure}. Nearest-neighbor positions are given by the three vectors
\begin{align}
\mathbf{T}_1 &= a \begin{bmatrix} 1 \\ 0 \end{bmatrix}, \quad
\mathbf{T}_2 = -\frac{a}{2} \begin{bmatrix} 1 \\ \sqrt{3} \end{bmatrix}, \quad
\mathbf{T}_3 = \frac{a}{2} \begin{bmatrix} -1 \\ \sqrt{3} \end{bmatrix}
\end{align}
with $a \approx 0.33$ nm the lattice constant~\cite{di1973preparation}.


The conduction band in 1H-TaS$_2$ has a width on the order of 2 eV and consists of three, spin-split Fermi pockets---one hole pocket around the $\Gamma$ point and another two hole pockets around the $K$ and $K'$ points. The lack of inversion in the plane leads to a spin-orbit coupling of Ising type, which spin splits the bands. Note that the $z$ direction of the spin remains a good quantum number due to the system's mirror symmetry $z \mapsto -z$~\cite{Mattheiss1973}. 

The itinerant electrons in TaS$_2$ have $d_{xy},\,d_{x^2-y^2}$, and $d_{z^2}$ orbital character requiring, in principle, a three-band description~\cite{Ritschel2015, Liu2013three}. We instead use an effective single-orbital model, which reproduces the key features of the low-energy bandstructure~\cite{yu2025quantum}. Namely, we consider a Hamiltonian with nearest-neighbor and next-to-nearest-neighbor hopping on the triangular lattice. In reciprocal space, the model is given by
\begin{equation}\label{eq:H}
    \begin{aligned}
 H = \sum_{\mathbf{k}, \sigma}  (\varepsilon_{\mathbf k \sigma}-\mu)c^{\dagger}_{\mathbf{k} \sigma} c^{\pdag}_{\mathbf{k} \sigma}
\end{aligned}
\end{equation}
with the dispersion
\begin{multline}
    \varepsilon_{\mathbf{k}\sigma} =  \sum_{\alpha=1}^3 \left[ -t_{\rm NN}\cos(\mathbf{T}_\alpha \cdot \mathbf{k}) + t_{\rm NNN}\cos(\mathbf{T}'_\alpha \cdot \mathbf{k})\right. \\ 
    + \left.\sigma\,\lambda  \sin(\mathbf{T}_\alpha \cdot \mathbf{k}) \right].
\end{multline}
In the following, we use $t_{\rm NN}=0.09$ eV for the hopping matrix element between nearest neighbors, $t_{\rm NNN}=0.26$ eV for the hopping element between next-nearest neighbors. Further, $\mathbf{T}_\alpha' = 2\mathbf{T}_{\alpha} + \mathbf{T}_{\alpha+1}$ connects next-nearest neighbors, $\lambda$ is the (Ising) spin-orbit-coupling strength between nearest-neighbor sites, and $\mu$ is the chemical potential.
The band structure of this model for $\lambda = 0.28$ eV is shown in Fig.~\ref{fig:H-structure}(b). While these parameters are set to reproduce the known band structure~\cite{de2018tuning,Nayak2021}, we consider below the dependence of our results on $\lambda$ and will therefore use it as a tuning parameter. 

The model defined in Eq.~\eqref{eq:H} undergoes two Lifshitz transitions as a function of the dimensionless spin–orbit-coupling strength $\lambda/t_{\rm NN}$~\cite{Wang2024}. For fixed chemical potential and varying spin-orbit-coupling-strength, the two transitions almost coincide at $\lambda/t_{\rm NN}\approx 2.6$.

\subsection{Disorder potential}
We turn now to the part of the Hamiltonian describing the elastic scatterers:
\begin{align}\label{eq:L_imp}
    H_{\rm imp} = \frac{1}{\Omega}\sum_{j=1}^{N_{imp}}\sum_{\mathbf{k},\mathbf{p}}V_{\mathbf{p},\mathbf{k}} e^{i\mathbf{R}_j\cdot (\mathbf{k} -\mathbf{p})}c^\dagger_{\mathbf{k}}c^{\pdag}_{\mathbf{p}}.
\end{align}
Here, $V_{\mathbf{p},\mathbf{k}}$ is the scattering potential in momentum space, $\mathbf{R}_j$ are the locations of the impurities and $\Omega$ is the system's volume. We consider scatterers of two different types:\\
(1) Impurities located on Ta sites, for which the respective scattering can be described by a point defect in real space. Such a scatterer can be modeled using a delta function in space and, hence,
\begin{align}\label{eq:Sd_point}
   V_{\mathbf{p},\mathbf{k}} = V_0 .
\end{align}\\
(2) Impurities on the S sites with
\begin{align}\label{eq:Sd_extended}
   V_{\mathbf{p},\mathbf{k}} = V_1\left[1+e^{i(\mathbf{k-p})\cdot\mathbf{T}_1} +e^{-i(\mathbf{k-p})\cdot\mathbf{T}_2}\right]. 
\end{align}
To understand the origin of this term, we recall that the crystal structure of the 1H layer is trigonal prismatic, such that each chalcogen defect equally affects three transition metal sites surrounding its location, as depicted in Fig ~\ref{fig:H-structure}(a). Note that these chalcogen defects are abundant in TMDs, due to vacancies or ad-atoms. For example, in the 4Hb samples they result from S-Se substitution~\cite{Ribak2020}.

\section{Abrikosov-Gorkov Theory}
In what follows, we review the essential steps in the generalization of AG theory~\cite{AAAbrikosov1959} to the model with Ising SOC and $D_{3h}$ symmetry, following Ref.~\cite{Dentelski2020}. We compute $\tau(\mathbf k)$, the single-particle lifetime taken from the electronic self-energy, and $\Gamma$, a disorder-induced cutoff scale extracted from the dressed superconducting susceptibility within the Cooperon approximation.

Before starting, it is worth noting that according to AG theory, a conventional superconductor subject to short-ranged and non-magnetic disorder is protected against pair breaking, such that the pair-breaking rate vanishes, $\Gamma=0$, despite a finite single-particle scattering rate $1/\tau \neq 0$. In contrast, in the simplest single-band treatment of unconventional superconductors, AG theory predicts a finite pair-breaking effect of order $\tau \Gamma \sim 1$.

\begin{figure}
    \centering
    \includegraphics[width=1\linewidth]{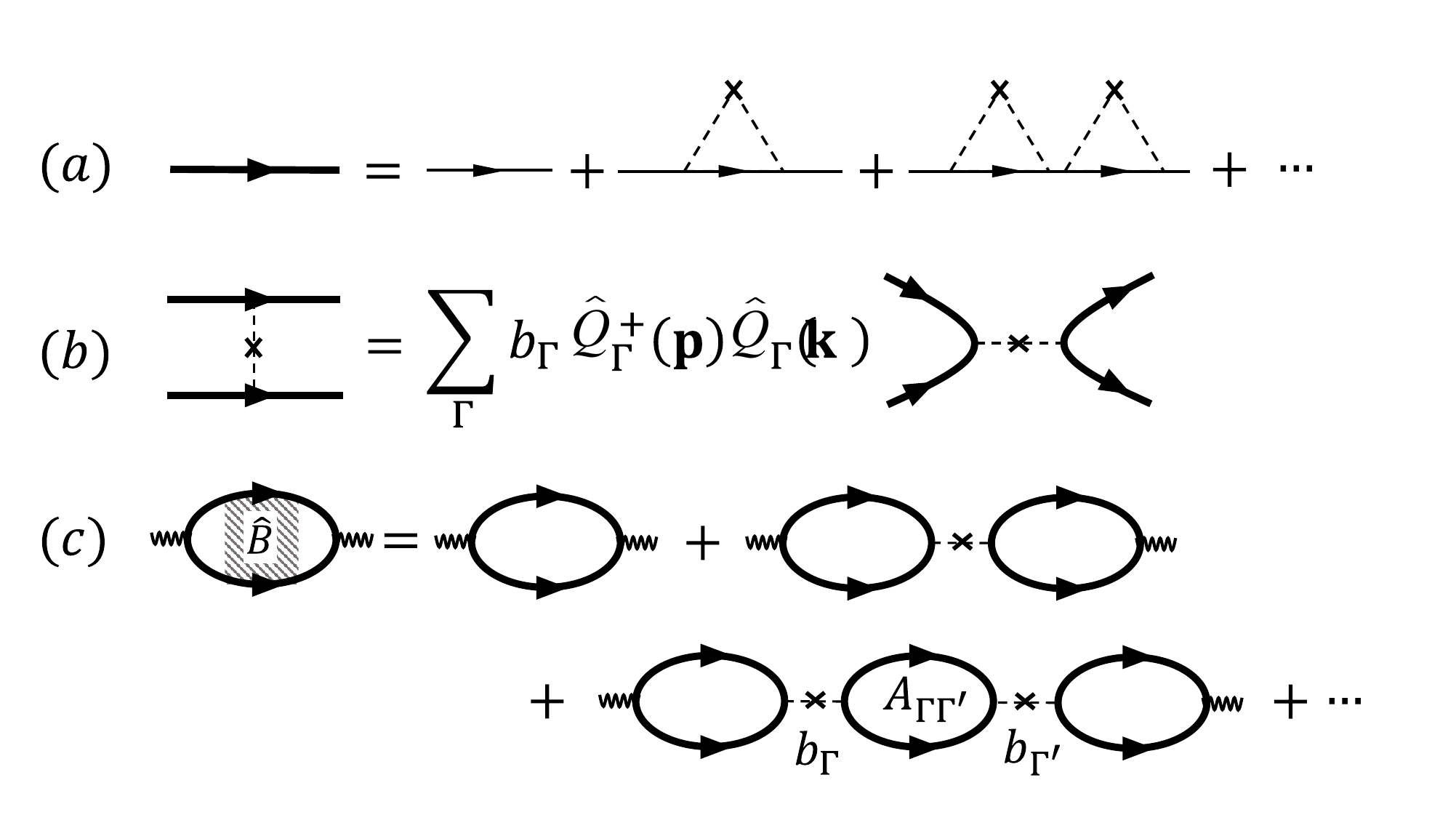}
 \caption{ (a) The sum of Feynman diagrams leading to Eq.\eqref{eq.dyson}. The first non-trivial diagram (the tent) corresponds to the self-energy Eq.~\eqref{eq:self_energy}. (b) The diagram show how we change basis from particle-hole to particle-particle and decomposes into scattering channels of the orthogonal basis functions $\hat Q_\Gamma(\textbf{k})$. (c) The diagram depicts the scattering of two particles on an impurity and the scheme of all these processes. The r.h.s of the equation denotes the dressed susceptibility $B$~\ref{eq:B}, which is obtained by connecting an infinite series of bare susceptibilities $A$ (Eq.~\ref{eq:A}) with a pair scattering vertex.
 }
    \label{fig:Diagram}
\end{figure}

\subsection{Single-particle scattering rate} 
{  In this section we perform the first step in dressing the pairing vertex with disorder: We calculate the single-particle lifetime resulting from the disorder Hamiltonian Eq.~\eqref{eq:L_imp} within the first Born approximation.  } We compute the electronic self-energy by summing up the Dyson series after averaging over disorder,
\begin{align} \label{eq.dyson} 
    \mathcal{G}^{-1}_\sigma(\mathbf{k}, i\omega_n)={G_\sigma^{-1}(\mathbf{k}, i\omega_n)-\Sigma_{\sigma}(\mathbf{k}, i\omega_n)}\,,
\end{align}
where the bare Green's function is given by 
\begin{align}\label{eq:bareG}
     G^{-1}_\sigma(\mathbf{k}, i\omega_n) = {i\omega_n - \varepsilon_{\mathbf{k}\sigma} + \mu}\,,
\end{align}
and $i\omega_n = 2\pi T(n+1/2)$ are Fermionic Matsubara frequencies. Note that the Green's function $\hat{\mathcal{G}}(\mathbf{k}, i\omega_n)$ is a diagonal $2\times 2$ matrix here.

As long as the Fermi energy is much greater than the scattering rate (or equivalently, the Fermi wavelength is much smaller than the mean-free path $k_{\rm F} \ell \gg 1$), the leading contribution is the ``tent'' diagram shown in Fig~\ref{fig:Diagram}(a), which yields
\begin{equation}
 \begin{aligned}\label{eq:self_energy}  
    \Sigma_{\sigma}(\mathbf{k},i\omega_n) = \frac{N_{\rm imp}}{ \Omega^2}\sum_\mathbf{p} V_{\mathbf{k}, \mathbf{p}}G_{\sigma}(i\omega_n,\mathbf{p})V_{\mathbf{p},\mathbf{k}}.
\end{aligned}
\end{equation}
Note that we absorbed the trivial first-order term, $V_{\mathbf{k}, \mathbf{k}}$, into the chemical potential in Eq.~\eqref{eq:H}. 

In the above, we neglect vertex corrections and restrict ourselves to rainbow diagrams. 
This approximation is controlled provided that impurity scattering remains weak compared to the relevant electronic energy scales. 
To ensure parametric control, we require that the disorder-induced broadening remains small compared to the characteristic band-structure scale, 
\begin{equation}
    \uptau\,\epsilon_0 \gg1 ,
\end{equation}
where \( \epsilon_0 \) denotes the minimal energy separation between the Fermi level and nearby extremal points of the dispersion.

{  The dispersion of the Hamiltonian Eq.~\eqref{eq:H} has several van Hove singularities relatively close to the Fermi energy at half-filling. These mark the transition from hole like Fermi surfaces around $K$ and $\Gamma$ and particle like Fermi surfaces around $M$ (the so called ``dog bone'' surfaces). For the parameters we use, the smallest value of $\uptau \epsilon_0$ is 6, which is obtained for the strongest disorder potential considered and $\lambda = 2 t_{\textrm{NN}}$. Note that ARPES measurements~\cite{almoalem2024charge} suggest that the Fermi energy is even closer in 4Hb-TaS$_2$, such that  disorder effects beyond the first Born approximation may become relevant. } 

To proceed, we consider a concrete example, the case of extended defects, Eq.~\eqref{eq:Sd_extended}. 
In this case, the impurity form factor in Eq.~\eqref{eq:self_energy} is given by 
\begin{equation}
     \begin{aligned}
     V_{\mathbf{k}, \mathbf{p}}V_{\mathbf{p},\mathbf{k}} = V_1^2\Big\{3 + 2\sum_{n} \cos[(\mathbf{k} - \mathbf{p})\cdot \mathbf{T}_n]\Big\}.
\end{aligned}
 \end{equation}
It is convenient to decompose this form factor into irreducible representations (irreps) of the point group $D_{3h}$, given in Table~\ref{tab:irreps}. Concretely, we write 
\begin{align}\label{eq:Vexpanded}
    &V_{\mathbf{k},\mathbf{p}}V_{\mathbf{p},\mathbf{k}} = V_1^2\sum_{\Gamma} z _{\Gamma}\psi_\Gamma (\mathbf{k})\psi_{\Gamma}^*(\mathbf{p}),  
\end{align}
where $\Gamma$ labels basis functions of the irreps of $D_{3h}$ and $z_\Gamma$ are their weights. 
Combining Eqs.~\eqref{eq:bareG}, \eqref{eq:self_energy}, and \eqref{eq:Vexpanded}, we find
 \begin{align} \label{eq:result-self-energy}
  &\Sigma_{\sigma}^\Gamma(i\omega_n) =   {2\pi z_{\Gamma}\over \nu_0\tau_0\Omega} \sum_{\mathbf{p}} \frac{\psi_{\Gamma}^*(\mathbf{p})}{i\omega_n - \varepsilon_{\mathbf{p}\sigma} + \mu}\\
  & \Sigma_\sigma(\mathbf{k},i\omega_n) =  \sum_{\Gamma}\psi_{\Gamma}(\mathbf{k})\Sigma_{\sigma}^\Gamma(i\omega_n)\,,
\end{align}
where $\tau_0  =1/ (2\pi  \nu_0 n_{\rm imp}V_{1}^2)$ is the time scale that reflects the scattering strength of the impurity potential,  $n_{\rm imp} = N_{\rm imp}/\Omega$ and $\nu_0$ is the average density of states at the Fermi level. 
The (momentum-dependent) quasiparticle scattering rate is then obtained from
\begin{align} \label{eq:life-time}
    \mathrm{Im} \Sigma_\sigma(\mathbf{k}, i\omega_n)
     = -\frac{\text{sign}(\omega_n)}{2\tau_\sigma(\mathbf{k})}.
\end{align}

\begin{table}[t]
\caption{\label{tab:irreps}
Basis functions $\Psi_\Gamma(\mathbf{k})$ for the irreducible representations (irreps) of D$_{3h}$ with their corresponding irrep in $D_{6h}$ in parentheses with $\phi_\pm = \exp\left(\pm\frac{2\pi i}{3}\right)$. Note that for the pairing channels, $\hat Q_\Gamma(\mathbf{k})$, we multiply by $(-i\sigma^y)/\sqrt{2}$ (spin singlet) and $\sigma^x/\sqrt{2}$ (spin triplet).}
\begin{tabular}{c|cl}
$D_{3h}$ & Basis function, $\Psi_{\Gamma}(\vk)$ & \\
\hline
\multirow{3}{*}{$A_{1}'$} & $1$ & ($A_{1g}$)\\ 
 & \; $e_\mathbf{k}(\mathbf{k}) \equiv \sum_{\alpha} \cos(\mathbf{k} \cdot \mathbf{T}_\alpha)$& ($A_{1g}$)\\ 
 & \;$o_\mathbf{k}(\mathbf{k}) \equiv \sum_{\alpha} \sin(\mathbf{k} \cdot \mathbf{T}_\alpha)$ & ($B_{1u}$) \\ 
\hline
\multirow{2}{*}{$E'$} & $e^\pm_\mathbf{k} \equiv \sum_{\alpha} \phi_{\pm}^n \cos(\mathbf{k} \cdot \mathbf{T}_\alpha)$  & ($E_{2g}$) \\
 & $o^\pm_\mathbf{k} \equiv \sum_{\alpha} \phi_{\pm}^n \sin(\mathbf{k} \cdot \mathbf{T}_\alpha)$ & ($E_{1u}$) \textbf{}\\
\end{tabular}
\end{table}

\subsection{Pair-Breaking Rate}
We now turn to the second step, the computation of the dressed pairing susceptibility. We start with the bare, frequency-resolved pairing susceptibility,
\begin{align}
   &A_{\Gamma,\Gamma'}(\omega_n) =\label{eq:A} \\&{1\over \Omega}\sum_{\mathbf{k}} \mathrm{Tr}\left[\hat{\mathcal{G}}(\mathbf{k},i\omega_n)\hat{Q}^\dagger_\Gamma(\mathbf{k})\hat{\mathcal{G}}^T(-\mathbf{k},-i\omega_n)\hat{Q}_{\Gamma'}(\mathbf{k})\right],\nonumber 
\end{align}
where $\hat{Q}_\Gamma$ corresponds to the pairing channel of irrep $\Gamma$, see Table \ref{tab:irreps}.

Next, we dress the susceptibility with impurity scattering via the Cooperon ladder. Rewriting the scattering vertex in the particle-particle channel,
\begin{align}\label{eq:b}
    V_{\mathbf{p},\mathbf{k}}V_{ -\mathbf{p},-\mathbf{k} }\delta_{\alpha \beta}\delta_{\gamma \delta} = \sum_\Gamma b_\Gamma [\hat{Q}_\Gamma ^\dagger (\mathbf{p})]_{\alpha \gamma} [\hat{Q}_\Gamma  (\mathbf{k})]_{\delta \beta},
\end{align}
we obtain {  the dressed frequency-resolved pairing susceptibility}
\begin{equation}\label{eq:B}
    \hat{\mathcal{B}}(i\omega_n)= (\hat{I}-\hat{A}(i\omega_n) \hat{b})^{-1} \hat{A}(i\omega_n).
\end{equation}
{  Finally, the pairing susceptibility matrix is obtained through a Matubara sum 
\begin{equation}\label{eq:chi}
\hat \chi(T) = T \sum_{\omega_n}\hat{\mathcal B}(i\omega_n).
\end{equation}}

The matrix $\hat{\mathcal{B}}(i\omega_n)$ is block-diagonal in the irreducible representations $A_1'$ and $E'$. We project the disorder potential, self-energy, and susceptibilities onto five basis functions (see Table~\ref{tab:irreps}), two of which transform under two-dimensional irreducible representations. Accordingly, $\hat{\mathcal{B}}$ decomposes into a $3 \times 3$ block spanned by $\{1, e(\mathbf{k}), o(\mathbf{k})\}$ and a $2 \times 2$ block spanned by $\{e_\pm(\mathbf{k}), o_\pm(\mathbf{k})\}$. Diagonalizing each block yields eigenmodes $\eta$ of the disorder-dressed susceptibility.

We label the corresponding eigenvalues as $\{A_1'^{(1)}, A_1'^{(2)}, A_1'^{(3)}\}$ for the $A_1'$ irrep and $\{E'^{(1)}, E'^{(2)}\}$ for the $E'$ irrep. Notably, $A_1'^{(1)}$ is dominated by the basis function $1$ and therefore corresponds most closely to an $s$-wave channel.

The corresponding eigenvalues exhibit the low-frequency asymptotic form
\begin{equation}\label{eq:B_eta}
\mathcal B_\eta(i\omega_n) \sim \frac{\pi \nu_\eta}{|\omega_n| + \Gamma_\eta/2}, \quad \omega_n \to 0,
\end{equation}
which defines the scale $\Gamma_\eta$. Here, $\nu_\eta$ has units of density of states and encodes the channel-dependent response of the bare electron gas.  

\begin{figure}
    
    \centering
    \includegraphics[width=0.8\linewidth]{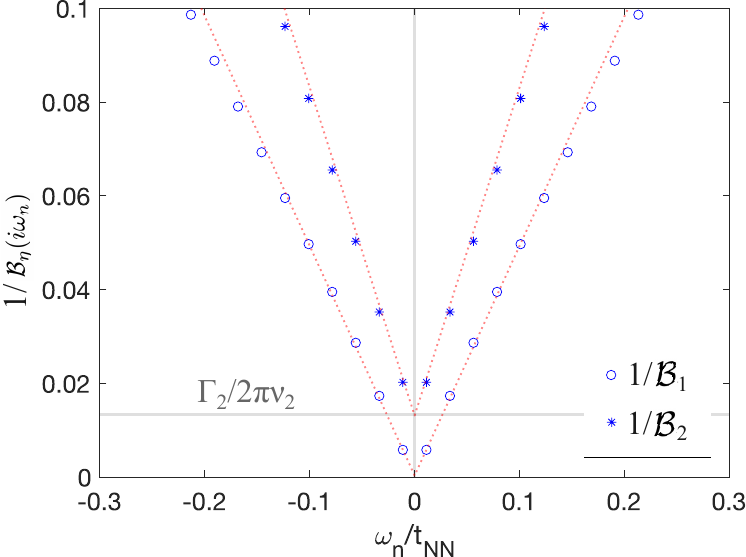}
    \caption{ Two of the eigenvalues of the susceptibility matrix $\hat{\mathcal B}(i\omega_n)$, $B_1$ and $B_2$ vs. Matsubara frequency $\omega_n$ for $\tau_0 t_{\rm NN} = 13$ and for a point defect potential Eq.~\eqref{eq:Sd_point}. The red dotted lines are the fits from which the parameter $\Gamma_1$ and $\Gamma_2$ are extracted. As can be seen, the extrapolation of $1/\mathcal B_1$ channel to $\omega_n\to 0$ approaches zero implying $\Gamma_1 \ll 1$, which is consistent with Anderson's theorem. On the other hand, $1/\mathcal B_2$ extrapolates to a finite  value, very close to $1/2\pi \nu_2 \tau$, which is consistent with AG theory.  (The units of the $y$-axis, $1/\mathcal B_\eta$, $t_{\rm NNN}^2a^2$, where $a$ is the lattice constant.)}
    \label{fig:fit}
\end{figure}

Figure~\ref{fig:fit} shows the inverse of the eigenvalues of the matrix $\hat{\mathcal{B}}$ vs. Matsubara frequency for the channels $A_1'^{(1)}$ (predominantly ``s-wave'') and $A_1'^{(2)}$ (predominantly ``f-wave'') for the parameters in Fig.~\ref{fig:H-structure}, $\tau_0 t_{\rm{NN}} = 26$, and for point-defect disorder. For this purpose, we compute the self-energy, bare susceptibility, and Cooperon numerically using Monkhorst-Pack integration~\cite{Monkhorst1976}. The parameter $\Gamma_\eta$ is extracted from the low-frequency behavior of eigenvalues Eq.~\eqref{eq:B_eta}. 
As can be seen, $1/\mathcal{B}_2$ intercepts the y-axis at a finite value that allows to extract $\Gamma_2$. In contrast, $1/\mathcal{B}_1$ intercepts at zero, implying $\Gamma_1 = 0$ in accord with Anderson's theorem and AG theory. 

This procedure is reliable provided the Fermi-surface-based approximations underlying the AG framework remain valid. In the vicinity of Lifshitz transitions or van Hove singularities, these assumptions break down, and the extracted $\Gamma_\eta$ may become unreliable or even unphysical.

{ 
\subsection{Computation of the transition temperature}\label{sec:transition_temp}

The quantities $\Gamma_\eta$ introduced above should be understood as disorder-induced cutoff scales associated with the eigenmodes of the pairing susceptibility. They are properties of the normal-state Hamiltonian and disorder potential alone, and therefore do not require specifying the pairing interaction. Consequently, $\Gamma_\eta$ should not, in general, be interpreted as universal pair-breaking rates governing the suppression of the superconducting transition temperature.

To determine $T_{\rm c}$, one must additionally specify the pairing interaction
\begin{equation}
H_I = \frac{1}{\Omega}\sum_{\mathbf k , \mathbf p}
U_{\Gamma \Gamma'}
\,c^\dagger_{\mathbf p} \hat Q_{\Gamma}(\mathbf p)c^\dagger_{-\mathbf p}\;
c_{-\mathbf k} \hat Q_{\Gamma'}(\mathbf k)c_{\mathbf k}.
\end{equation}
The superconducting instability is then determined by the interaction matrix above and the pairing susceptibility Eq.~\eqref{eq:chi},
\begin{equation}
    \det\!\left[\hat I - \hat \chi(T)\hat U\right] = 0.
\end{equation}

In general, the interaction matrix $\hat U$ and the susceptibility matrix $\hat\chi$ need not be simultaneously diagonalizable. In this case, the superconducting instability is determined by the full matrix equation above, and the transition temperature cannot be expressed solely in terms of the quantities $\Gamma_\eta$. Nevertheless, the cutoff scales $\Gamma_\eta$ remain physically meaningful even in this
generic case. As shown in Appendix~\ref{app:theorem}, they provide an interaction-independent upper bound on the superconducting instability.

In the special case where $\hat U$ and $\hat\chi$ share the same eigenbasis, the pairing channels decouple and one recovers an Abrikosov--Gor'kov equation for each eigenmode,
\begin{equation}\label{eq:AG_Tc}
\log \frac{T_{c,\eta}}{T_{c,\eta}^{0}}
=
\Psi\!\left(\frac12\right)
-
\Psi\!\left(
\frac12+\frac{\Gamma_\eta}{4\pi T_{c,\eta}}
\right),
\end{equation}
where
\[
T_{c,\eta}^{0}
=
\frac{2\omega_D}{\pi}
\exp\!\left[-\frac{1}{\nu_\eta U_\eta}+\gamma\right]
\]
is the transition temperature in the clean limit.

Throughout this work we therefore focus on the interaction-independent quantities $\Gamma_\eta$, which characterize the disorder-induced suppression of the eigenmodes of the pairing susceptibility. Whenever the pairing interaction approximately shares the same eigenbasis as the susceptibility, these quantities directly determine the suppression of $T_{\rm c}$ through Eq.~\eqref{eq:AG_Tc}; otherwise, the full matrix equation above must be solved.
}

\subsection{Momentum-relaxation rates}
As discussed in the introduction, experiments on 4Hb-TaS$_2$ present the following puzzle: how can signatures of non-$s$-wave superconductivity persist in a material that appears to lie in the dirty limit ($\xi > \ell$)? This reasoning relies on identifying the superconducting pair-breaking rate with the momentum relaxation rate inferred from transport measurements. Our objective is to test this assumption within a microscopic framework incorporating a realistic band structure and impurity model. Accordingly, in parallel to our computation of the pair-breaking rate  we also compute the transport momentum relaxation rate within the same model and under the same set of assumptions. In the following, we outline the methodology used to evaluate this rate.

We employ Boltzmann’s transport equation using the single-life-time approximation, which yields
\begin{align}\label{eq:conductivity}
    \sigma_{\alpha \beta} = {1\over \Omega}\sum_{\mathbf k\sigma} \; \uptau_{\text{tr}}^\sigma (\mathbf{k})v_{\mathbf{k}\sigma, \alpha}v_{\mathbf{k}\sigma,\beta} \left( -\frac{df(\varepsilon)}{d\varepsilon}\right)_{\varepsilon = \varepsilon_\mathbf{k\sigma}} ,
\end{align}
where $v_{\mathbf{k}\sigma, \alpha} = {\partial \varepsilon_{\mathbf{k}\sigma}} /{\partial {{k_\alpha}}}$,
\begin{equation}
\begin{aligned}
    \frac{1}{\uptau_{\text{tr}}^\sigma(\mathbf{k})} = {1\over \Omega} \sum_{\mathbf p}\,V_{\mathbf{k,p}}V_{\mathbf{p,k}} 
    \left(1-\hat{\boldsymbol{v}}_{\mathbf{p}\sigma}\cdot \hat{\boldsymbol{v}}_{\mathbf{k}\sigma} \right)\delta (\epsilon_{\mathbf{p}\sigma}-\epsilon_{\mathbf{k}\sigma}),\label{eq:tau_D}
\end{aligned}     
\end{equation}
and $\hat{\boldsymbol{v}}_{\mathbf{k}\sigma} = \boldsymbol{v}_{\mathbf{k}\sigma}/{v}_{\mathbf{k}\sigma}$.
To make contact with experiments, we convert the conductivity to a single typical time scale defined as 
\begin{align}
   \uptau_{\rm D}  = {\sigma_{xx}\over \sigma_0 k_{\rm F} v_{\rm F} },
\end{align}
where $\sigma_{xx}$ is computed from Eq.~\eqref{eq:conductivity}, $\sigma_0 = 2e^2/h$ is the quantum of conductance, and $v_{\rm F} = \langle |\boldsymbol{v}_{\mathbf{k}\sigma}| \rangle_{\rm FS_\sigma}$  is the  Fermi velocity averaged over the Fermi surface.

\section{Results}
\begin{figure} \centering \includegraphics[width=0.75\linewidth]{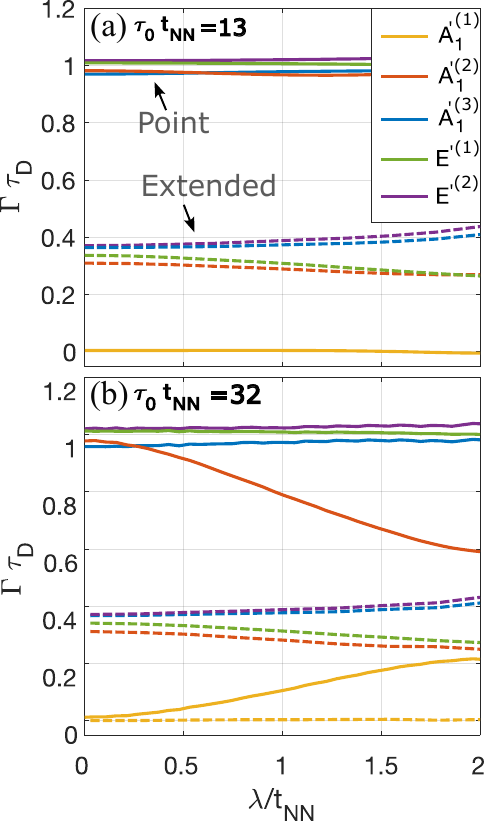} \caption{(a) The product $\uptau_{\rm D} \Gamma$ vs. $\lambda/t_{\rm NN}${ , the dimensionless spin–orbit-coupling strength,} for point defect Eq.~\eqref{eq:Sd_point} (solid) and extended defect  Eq.~\eqref{eq:Sd_extended} (dashed lines) for strong disorder $\tau_0t_{\rm{NN}} = 13$. The channels $A_1'^{(1,2,3)}$ and $E'^{(1,2)}$ correspond to different eigenvalues of the susceptibility Eq.~\eqref{eq:B_eta} belonging to the irreps $A_1'$ and $E'$, respectively. (b) The same as (a) for weaker disorder $\uptau_0 t_{\rm NN} = 32$. The color coding is the same as for panel (a) for both extended and point defects. } \label{fig:main result} \end{figure}

We now turn to discuss the results of the calculations, which culminate in the parameter $\Gamma_\eta$ extracted from the asymptotic behavior, Eq.~\eqref{eq:B_eta}, as shown in Fig.~\ref{fig:fit}. We begin with the case of point-like disorder, $V_{\mathbf k,\mathbf p} = V_0$, Eq.~\eqref{eq:Sd_point}, focusing first on the strong-disorder limit $\uptau_0 t_{\rm NN} = 13$, shown in Fig.~\ref{fig:main result}(a).

In this limit, the results are in good agreement with AG theory. The trivial ``s-wave'' channel, labeled $A_1'^{(1)}$, remains protected with $\uptau_{\rm D}\Gamma = 0$, while all other channels exhibit $\uptau_{\rm D}\Gamma \approx 1$. These values depend only weakly on the specific channel and on the strength of the spin-orbit coupling $\lambda$, indicating that the basic AG phenomenology remains robust in this regime.

Next, we turn to the weaker-disorder case, $\uptau_0 t_{\rm NN} = 32$, shown by the solid lines in Fig.~\ref{fig:main result}(b). The overall behavior remains qualitatively similar: the $A_1'^{(1)}$ channel continues to exhibit a strongly suppressed pair-breaking rate, while the remaining channels have $\uptau_{\rm D}\Gamma$ of order unity. However, in contrast to the strong disorder limit, a more pronounced dependence on $\lambda$ develops, signaling deviations from the simplest AG expectations. In particular, the pair-breaking rate of the $A_1'^{(2)}$ channel is reduced at the expense of $A_1'^{(1)}$. By contrast, the pair-breaking rates of the two-dimensional irreps $E'$ remain largely unaffected and do not exhibit any significant suppression.

The main result of this work is revealed upon comparing these results to the case of extended disorder, Eq.~\eqref{eq:Sd_extended}, shown by the dashed lines in Fig.~\ref{fig:main result}(b). In this case, we observe a pronounced suppression of the pair-breaking rate compared to the transport scattering rate $1/\uptau_{\rm D}$. In particular, at $\lambda = 0$ we find $\uptau_{\rm D}\Gamma \approx 1/3$ for all unconventional pairing channels, substantially smaller than the value $\uptau_{\rm D}\Gamma \sim 1$ expected from AG theory. As $\lambda$ increases, the pair-breaking rates split: $A_1'^{(2)}$ and $E'^{(1)}$ decrease, while $A_1'^{(3)}$ and $E'^{(2)}$ increase.

Thus, increasing $\lambda$ reduces the pair-breaking rate for one of the  the multi-dimensional irrep $E'$. To analyze its structure, we examine the weights of the corresponding basis $\{e_\pm (\mathbf k),o_\mp(\mathbf k) \}$ (see Table~\ref{tab:irreps}). Specifically, we parametrize the members of the $E'^{(1)}$ channel as
\begin{equation} \label{eq:F2} 
    \hat{Q}_{E'^{(1)}}^\pm(\mathbf k) = \delta \,e_\pm(\mathbf k)(-i\hat{\sigma}^y)+\kappa\,o_\mp(\mathbf k)\, \hat{\sigma}^x\,,
\end{equation}
where $\delta$ and $\kappa$ denote the respective weights. We find that the channel exhibiting reduced pair breaking is dominated by the odd-parity component $o_\mp(\mathbf k)$, i.e., $|\kappa|^2 > |\delta|^2$. This indicates that extended defects preferentially preserve pairing states with this internal structure. For example, at $\lambda/t_{\mathrm NN} = 1.43$ and $\tau_0 t_{\mathrm NN} = 32$, we find $|\delta|^2 \approx 0.25$ and $|\kappa|^2 \approx 0.75$.
{  Thus, we find a reduced pair-breaking rate as compared to AG theory by a factor of the order of 1/3.  If we focus on the two-dimensional irreps $E'$ and consider the splitting by the spin-orbit coupling, our calculations suggest that disorder is least detrimental for the odd pairing state $e_\pm (\mathbf k)$ and as such, this pairing state would be the most consistent $E'$ state given current experiments.}

This result demonstrates that, for realistic extended impurity potentials, the pair-breaking scale $\Gamma$ can be strongly decoupled from the momentum relaxation rate. Physically, this suppression originates from the nontrivial momentum structure of the disorder potential, which partially matches the internal structure of the superconducting gap. As a result, scattering processes that efficiently relax momentum do not necessarily mix pairing states with opposite signs, and are therefore less effective at breaking Cooper pairs. This provides a natural mechanism by which unconventional superconductivity can remain robust even in systems with relatively short transport lifetimes.

To further elucidate the deviations from AG behavior already visible in the point-defect case at weaker disorder, we analyze the internal structure of the $A_1'^{(1)}$ channel by decomposing it in the basis $\{1, e(\mathbf k), o(\mathbf k)\}$,
\begin{equation} \label{eq:F1} 
    \hat{Q}_{A_1'^{(1)}}(\mathbf k) = [\alpha + \beta \,e(\mathbf k)](-i\hat{\sigma}^y)+\gamma\,o(\mathbf k)\, \hat{\sigma}^x\,.
\end{equation}

\begin{figure} \centering \includegraphics[width=1\linewidth]{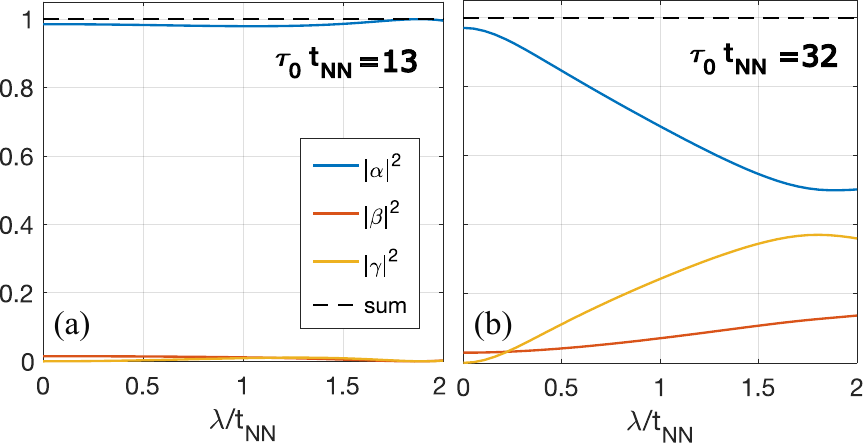} \caption{The weights of the channel $A_1'^{(1)}$, Eq.~\eqref{eq:F1}, closest to the `s-wave'', where $\alpha$, $\beta$ and $\gamma$ are the coefficient of the functions $1$, $e(\mathbf k)$ and $o(\mathbf k)$ in Table \ref{tab:irreps}. Panel (a) corresponds to strong disorder $\tau_0 t_{\mathrm NN} = 13$ and panel (b) to weak disorder $\tau_0 t_{\mathrm NN} = 32$.} \label{fig:weights} \end{figure}

The corresponding weights are shown in Fig.~\ref{fig:weights}. In the strong-disorder limit, the non-trivial components $e(\mathbf k)$ and $o(\mathbf k)$ are suppressed, and the pairing channel is dominated by the uniform basis function $1$. In contrast, for weaker disorder the extended components $\beta$ and $\gamma$ become sizable and grow with $\lambda$. This leads to mixing between the uniform and extended basis functions, generating a finite pair-breaking rate in the nominally ``s-wave'' channel $A_1'^{(1)}$, accompanied by a corresponding reduction in $\Gamma$ for the neighboring channel $A_1'^{(2)}$.

We emphasize, however, that this mixing effect constitutes a secondary correction within the point-defect scenario. The dominant effect highlighted above is the parametrically reduced pair-breaking rate induced by extended disorder potentials.

\section{Conclusions}
We have computed the pairing susceptibility of 1H-type transition metal dichalcogenides in the presence of non-magnetic disorder, focusing on two physically relevant types of scatterers: point-like defects on the transition metal sites and extended defects associated with chalcogen substitution.

\begin{figure}
     \centering
     \includegraphics[width=0.9\linewidth]{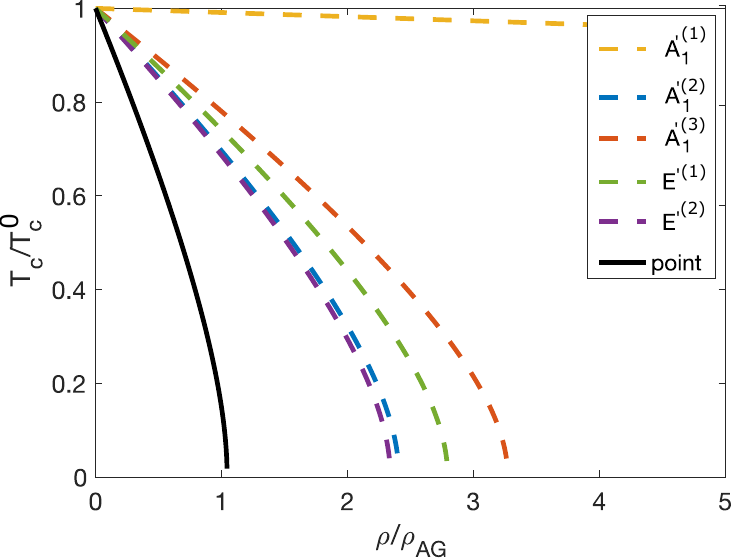}
     \caption{ Normalized critical temperature $T_{\rm c} / T_{\rm c}^{0}$ as a function of normalized resistivity $\rho / \rho_{\rm AG}$ from Eq.~\eqref{eq:conductivity} different pairing channels and for an extended defect potential Eq.~\eqref{eq:Sd_extended}. Here $T_{c}^0 = t_{\rm NN}/40$ and $\rho_{\rm AG}$ is defined to be the critical value of resistivity at which $T_{\rm c} = 0$ for a point defect disorder.}
    \label{fig:tc_vs_rho}
 \end{figure}

This problem is motivated by the growing body of experimental evidence for unconventional superconductivity in TMD materials despite relatively high resistivities ($10$--$100~\mu\Omega\,\mathrm{cm}$), most notably in 4Hb-TaS$_2$, which exhibits a range of anomalous superconducting signatures. Reconciling these observations with the expected fragility of unconventional pairing to disorder is therefore a central theoretical challenge.

Within a generalized Abrikosov--Gor'kov framework adapted to the multiband structure and symmetry of the system, we first organize the pairing susceptibility into irreducible representations of the $D_{3h}$ point group of a single 1H layer. Assuming that the interaction matrix commutes with the susceptibility, the superconducting instability develops in its eigenchannels, allowing for a direct characterization of disorder effects on different pairing states.

Our central result is that extended defect potentials lead to a parametrically reduced pair-breaking rate compared to the transport scattering rate. As a consequence, unconventional superconducting states remain significantly more robust than predicted by conventional AG theory. This effect is clearly demonstrated in Fig.~\ref{fig:tc_vs_rho}, where we compare the suppression of $T_{\rm c}$ as a function of normalized resistivity for point and extended defects at fixed $\uptau_{\rm D}$. While point-like disorder reproduces the standard AG behavior, extended defects yield a substantial enhancement of the critical disorder strength required to suppress superconductivity.

Physically, this enhancement originates from the momentum structure of the impurity potential, which can partially match the internal structure of the superconducting gap and thereby reduce pair-breaking processes. This mechanism, which in the case studied here is most pronounced for the odd parity $E'$ state $e_\pm (\mathbf k)$, provides a natural route to stabilizing unconventional pairing in realistic materials with nontrivial disorder profiles.

At the same time, we find that this mechanism alone is not sufficient to fully account for the exceptionally weak suppression of superconductivity observed in 4Hb-TaS$_2$. This suggests that additional ingredients beyond the present framework play an important role. In particular, 4Hb-TaS$_2$ exhibits substantial charge transfer between T and H layers, which shifts the chemical potential toward a van Hove singularity along the $\Gamma$--$K$ direction. In this regime, the assumptions underlying the AG approach may break down, raising the possibility that strong-coupling or non-perturbative disorder effects further enhance the robustness of unconventional pairing.

{ 
Another important simplification of the present work is the use of an effective single H-layer description. In the real material, the alternating H and T layers introduce additional low-energy degrees of freedom that may further modify the relation between momentum relaxation and pair breaking. In particular, hybridization between the two H layers competes with the Ising spin-orbit splitting and alters the spin texture of the Fermi surfaces, potentially opening additional backscattering channels. Moreover, impurities in the intervening T layers may induce interlayer scattering and couple to low-energy magnetic degrees of freedom that are absent from the present model. These effects, together with the charge-transfer physics discussed above, constitute natural extensions of the present framework and may further improve the agreement between theory and experiment.}

{  
A natural extension concerns the distinction between extended impurity potentials and spatially correlated disorder. Throughout the present work, we assume that impurity positions are statistically independent, such that the momentum dependence of the scattering originates entirely from the finite spatial extent of the single-impurity potential. By contrast, spatial correlations between impurity positions introduce an additional momentum dependence through the impurity structure factor, selectively enhancing or suppressing scattering at particular momentum transfers. Since pair breaking depends only on scattering processes connecting regions of the Fermi surface with different superconducting phases, such correlations may either further suppress or enhance pair breaking, depending on the nature of the disorder correlations~\cite{PhysRevB.111.184514}. It would therefore be interesting to investigate the combined effects of extended impurity potentials and correlated impurity distributions in future work.}

Our results establish that the apparent conflict between high resistivity and unconventional superconductivity in H-structured TMDs is substantially alleviated taking realistic disorder potentials within a tractable single-layer model into account, and they provide a concrete framework for systematically going beyond the standard AG paradigm in these systems.

\section{Acknowledgments}
We thank Amit Kanigel, Avraham Klein, Sofie Castro Holb\ae k and Aline Ramires for helpful discussions. J.R. acknowledges funding by the Simons Collaboration on “New Frontiers in Superconductivity” and ISF under grant No. 915/24.

\appendix
{ 

\section{Relation between the susceptibility eigenvalues and the superconducting instability}
\label{app:theorem}

In the main text, Section~\ref{sec:transition_temp},  we introduced the disorder-induced cutoff scales
$\Gamma_\eta$ through the low-frequency behavior of the eigenvalues of the
pairing susceptibility. In general, these quantities do not by themselves
determine the superconducting transition temperature, since the latter
depends also on the pairing interaction. In this appendix we show that,
even in the absence of this simplification, the susceptibility eigenvalues
provide an upper bound on the superconducting instability.

Let the linearized gap equation be
\begin{equation}
\det\!\left[\hat I-\hat\chi(T)\hat U\right]=0,
\end{equation}
where $\hat\chi(T)$ is the static pairing susceptibility and $\hat U$ is
the pairing interaction matrix in the space of irreps Table~\ref{tab:irreps}.

\noindent
\paragraph{Theorem.}
Let $\hat\chi$ be a positive-semidefinite Hermitian matrix and let
$\hat U$ be Hermitian.
Denote by $c_{\rm max}$ the largest eigenvalue of $\hat\chi$,
by $u_{\rm max}$ the largest eigenvalue of $\hat U$, and by
$\lambda_{\rm max}$ the largest eigenvalue of their product, $\hat U\hat\chi$.
Then

\begin{equation}
\lambda_{\rm max}
\le
u_{\rm max}\,c_{\rm max}.
\label{eq:bound}
\end{equation}

\noindent
\paragraph{Proof.}
Since $\hat\chi$ is positive semidefinite,
$\sqrt{\hat\chi}$ is well defined.
Furthermore, the matrices
$\hat U\hat\chi$ and
$\sqrt{\hat\chi}\,\hat U\,\sqrt{\hat\chi}$
have identical nonzero eigenvalues.
Using the Rayleigh-Ritz variational principle,
\begin{equation}
\lambda_{\rm max}
=
\max_{\|\vec v\|=1}
\vec v^\dagger
\sqrt{\hat\chi}\,
\hat U\,
\sqrt{\hat\chi}
\vec v .
\end{equation}
Now, using the fact that $\hat U$ is hermitian we can span the vector $\vec w=\sqrt{\hat \chi}\vec v$ using $\hat U$'s eigenbasis $\{(u_j,\vec e_j )\}$, which implies that the inner product  $w^\dagger \hat U w$ is maximized when $\vec w \parallel \vec e_{max}$. Therefore, 
\begin{align}
\lambda_{\rm max}
\le
u_{\rm max}
\max_{\|v\|=1}
v^\dagger
\hat\chi
v
\nonumber
=
u_{\rm max}c_{\rm max},
\end{align}
which proves Eq.~(\ref{eq:bound}).

Equation~(\ref{eq:bound}) demonstrates that the susceptibility eigenvalues
place an upper bound on the superconducting instability for any fixed
pairing interaction.
Consequently, although the cutoff scales $\Gamma_\eta$ extracted from the
susceptibility do not in general determine $T_c$, they nevertheless remain
interaction-independent measures of the sensitivity of different pairing
structures to disorder.
In the special case where $\hat U$ and $\hat\chi$ commute, the two matrices
are simultaneously diagonalizable, the bound becomes an equality for each
channel, and one recovers the Abrikosov--Gor'kov equation discussed in the
main text.}

\bibliography{export}

\end{document}